\documentclass[a4paper,11pt]{article}
\pdfoutput=1
\usepackage{jcappub} 
\usepackage[T1]{fontenc} 

\title{ Gravitational constant in multiple field  gravity }
\author{Habib Abedi}
\author{and Amir M. Abbassi}
\affiliation{Department of Physics, University of Tehran,\\
North Kargar Ave, Tehran-Iran.}
\emailAdd{h.abedi@ut.ac.ir} 
\emailAdd{amabasi@khayam.ut.ac.ir} 
\abstract{
In the present study, we consider general form of the Lagrangian $ f(R, \phi^{I}, X) $, that is a function of the Ricci scalar, multiple scalar fields and non-canonical kinetic terms. We obtain the effective Newton's constant deep inside the Hubble radius. 
We use Jordan and Einstein frames, and  study the conservation of energy-momentum tensor.
}
\begin{document}
\maketitle
\flushbottom
\section{Introduction}
\label{Introduction}
In 1998, supernova data suggested that the Universe is dominated by dark energy component  \cite{1}.
A simple model that fits with observations  includes the  positive cosmological constant.
The simplest dynamical model is proposed by a scalar field minimally coupled to gravity \cite{2,3}.
More complicated models are generally classified in scalar-tensor theories \cite{4b}.
Other possible models can be achieved by replacing  the Ricci scalar with a non-linear function of it \cite{5}. These models may be considered as generalized forms of scalar-tensor theories (Brans-Dicke) \cite{6}, where extra gravitational degree of freedom that comes from non-linearity of $ f(R) $, can be figured out as a scalar field.
Also other modified theories of gravity are tensor-vector-scalar theories (TeVes) \cite{7}, Conformal gravity \cite{8}, Lovelock gravity \cite{9}, Horava-Lifshitz \cite{10} and some teleparallel extensions of above theories, like $ f(T) $ gravity \cite{11}.
Multiple scalar field generalizations of single field models are also possible, that  could be non-minimally coupled.
The non-minimal couplings are necessary  in the renormalization  studying of scalar fields in curved  spacetime \cite{4}.
Non-minimal couplings can be removed by  convenient conformal transformations. 
In a conformal transformation one can go from non-minimal coupling  in Jordan frame to  Einstein frame with minimal coupling. But this transformation for systems with more than two degrees of freedom leads to non-canonical kinetic terms  in new frame, even if the old frame has only canonical kinetic terms \cite{12}.

In Einstein's theory of gravity, deep inside the Hubble radius, growth of matter perturbations  are governed by 
\begin{equation}
\delta^{''}+ \mathcal{H} \delta^{\prime} -  4 \pi G\rho \delta =0,
\end{equation} 
where $ \mathcal{H}=a H $ is comoving Hubble parameter, $ \delta $ is matter energy density contrast, $ G $ is Newtonian gravitational constant and prime denotes  derivation w.r.t conformal time \cite{Ellis}.
This equation for the other theories of gravity is modified by replacing effective Newton's constant.  
Many attempts have been made in obtaining this modification for f(R), scalar-tensor theory  and etc  \cite{14}, but it needs to be studied in its more general form. 
Our aim is to do this generalization.

In Section \ref{background Multi-field }, we extend field equations of multiple scalar fields to general form and obtain it for flat FLRW  (Friedmann-Lema\^{i}tre-Robertson-Walker) spacetime. Section \ref{perturbation in multifield model} is devoted to perturbed FLRW and to obtain effective Newton's constant of gravitation deep inside the Hubble radius. We  explicitly  calculate it for some specific models. Finally, last section is about conformal transformation between two frames of Jordan and Einstein, and we study the conservation of energy-momentum tensor. 
We use the signature $ (-,+,+,+) $ and notations $f_R:=\frac{\partial f}{\partial R} $, $ f_I:=\frac{\partial f}{\partial \phi^{I}} $, $ f_X:=\frac{\partial f}{\partial X} $, and $ \Gamma^L_{KJ} $ is the Levi-Civit\'{a} symbol, constructed from field space metric, $ \mathcal{G}_{IJ} (\phi^K) $. Also we  use  capital Latin  letters $I,J,K...$  to label $ N $ scalar fields and Greek  letters $ \alpha, \beta, ... $  for spacetime indices.
\section{Multiple fields in background spacetime}
\label{background Multi-field }
The general form of multiple scalar fields action with non-canonical kinetic terms can be written as  
\begin{equation}
S= \int dx^4 \sqrt{-g}\big( \frac{1}{2} f(R, \phi^I, X)+ \mathcal{L}_m \big), \label{action}
\end{equation}
where $ g $ is the determinant of metric $ g_{\mu \nu} $,
$ X:=-\frac{1}{2} \mathcal{G}_{IJ}(\phi^{K}) \nabla_{\mu}\phi^{I} \nabla^{\mu}\phi^{J} $ is the kinetic terms, $ \mathcal{G}_{IJ} $ is the field space metric, $ f(R, \phi^I, X) $ is a function of the Ricci scalar, scalar fields and kinetic terms and $\mathcal{L}_m$ is matter Lagrangian. We have  set $ 8 \pi G=1 $.
Variation of the action \eqref{action} w.r.t metric, yields
\begin{equation}
f_R G_{\mu\nu}+\frac{1}{2} g_{\mu\nu} ( f_R R-f ) + ( g_{\mu\nu} \square - \nabla_{\mu} \nabla_{\nu}) f_R - \frac{1}{2}  f_X \mathcal{G}_{IJ} \nabla_{\mu} \phi^{I} \nabla_{\nu} \phi^J = T_{\mu\nu}, \label{F1}
\end{equation}
where $ T_{\mu\nu}:= \frac{-2}{\sqrt{-g}} \frac{\delta  \mathcal{L}_m}{\delta g^{\mu \nu}}$ is the matter energy-momentum tensor and $ \square:=g^{\mu \nu} \nabla_{\nu} \nabla_{\mu} $. We dropped the arguments of $ \mathcal{G}_{IJ}(\phi^{K}) $. The generalized Klein-Gordon equation can be obtained by variation of the action \eqref{action} w.r.t scalar fields
\begin{equation}
\nabla_{\mu}\nabla^{\mu} \phi^{L}+\Gamma^{L}_{KJ}\nabla_{\mu}\phi^{K} \nabla^{\mu}\phi^{J}+\frac{\partial_{\mu}f_{X}}{f_{X}} \partial^{\mu}\phi^{L} +\frac{ \mathcal{G}^{IL}}{f_{X}}f_{I}=0. \label{KG}
\end{equation}
In the case of  $ \mathcal{G}_{IJ}=\delta_{IJ} $ and $ f_{R}=1 $, equations \eqref{F1} and \eqref{KG} reduce to canonical scalar field and gravity equations in general relativity. 
The metric of spatially flat FLRW  spacetime has the form: 
\begin{equation}
ds^2=a^2(\eta) [-d\eta^2+\delta_{ij}dx^i dx^j] \qquad i,j=1, 2, 3,
\end{equation}
where $ \eta $ is conformal time.
The energy-momentum tensor in perfect fluid form is:
\begin{equation}
T_{\mu \nu}=p g_{\mu \nu}+(p+\rho) u_{\mu} u_{\nu},
\end{equation}
where $ \rho $, $ p $ and $ u_{\mu} $ are the energy density, momentum and 4-velocity of the fluid with $ u_{\mu} u^{\mu}=-1 $, respectively.
The components (00) and (ii)  of field equation \eqref{F1} and scalar field equation \eqref{KG} become
\begin{eqnarray}
 -\frac{3\mathcal{H}^\prime}{a^2}f_{R}-\frac{1}{2} f-\frac{3\mathcal{H}}{a^2}f_{R}^{\prime}+\frac{1}{2a^2} f_X  \mathcal{G}_{IJ}\phi^{I \prime } \phi^{J \prime } &=& -\rho ,
\nonumber \\ \nonumber
-\frac{\mathcal{H}^\prime +2 \mathcal{H}^2}{a^2} f_{R}-\frac{1}{2} f -\frac{1}{a^2} f_{R}^{''} - \frac{2 \mathcal{H}}{a^2} f_{R}^{\prime} &=& p ,
\\ 
\phi^{'' L}+\Big(2  \mathcal{H} +\frac{f_{X}^{\prime}}{f_{X}} \Big) \phi^{L \prime }+\Gamma^{L}_{KJ} \phi^{K \prime } \phi^{J \prime } &=& a^2  \mathcal{G}^{IL} \frac{f_{I}}{f_{X}}.
\end{eqnarray}
The last equation can be written in compact form 
\begin{equation}
\mathcal{D}_{\eta}(a^2f_{X}\phi^{L \prime })=a^4 \mathcal{G}^{IL} f_{I},
\end{equation}
where we have used the definition of covariant derivative in field space for a vector $ A^{I} $,
\begin{equation}
\mathcal{D}_{J} A^{I}= \partial_{J} A^{I} + \Gamma^{I}_{KJ} A^{K}.
\end{equation}
Consequently $ \mathcal{D}_{\eta} \phi^{I \prime}:=\phi^{L \prime} \mathcal{D}_{L} \phi^{I \prime}= \phi^{I ''}+\Gamma^{I}_{KJ} \phi^{L \prime} \phi^{J \prime }$ is acceleration in field space. The operator $ \mathcal{D}_{\eta} $ acts as ordinary time derivative for quantities without any field space indices.
\section{Scalar perturbations in the multi-field theory}
\label{perturbation in multifield model}

The FLRW  spacetime perturbed in conformal 
Newtonian gauge, has the following form for scalar perturbations
\begin{equation}
ds^2=a^2(\eta) [-(1+2\varphi) d\eta^2+ (1-
2\psi)\delta_{ij} dx^i dx^j],
\end{equation}
where $ \varphi $ is modified Newtonian potential 
and $\psi $ is related to spatial curvature perturbation.
Scalar fields can be decomposed to background space independent $ \phi^{I}(\eta) $, and the perturbation  part, $  \delta \phi^{I}(\eta, x^{i}) $, that have both space and time dependence.
The components (00), (0i), (ij) with $i\neq j$ and (ii)  of the field equation \eqref{F1} and the equation \eqref{KG} in first order of perturbations become, respectively, 
\begin{eqnarray}
 -\delta\rho &=&-3\frac{\mathcal{H}^{\prime}}{a^2} \delta f_{R} +\frac{1}{a^2} \big[-3\psi^{''}+\nabla^{2}\varphi -3\mathcal{H}(\psi^{\prime}+\varphi^{\prime}) -6\mathcal{H}^{\prime} \varphi \big]  f_{R} -\frac{1}{2} \delta f  \nonumber
  \\ \nonumber &&+ \frac{ \mathcal{H} }{ a^2 } ( 2 \varphi - 4 \psi +\psi^{\prime} ) f_R^{\prime}+\frac{\nabla^2}{a^2} \delta f_R + \frac{1}{2 a^2} \delta f_X \mathcal{G}_{IJ}\phi^{I \prime }  \phi^{J \prime }
   \\ \nonumber &&+ \frac{1}{2a^2}    f_{X} \mathcal{G}_{IJ, K} \phi^{I \prime }  \phi^{J \prime } \delta \phi^{K} -\frac{\varphi}{a^2} f_{X} \mathcal{G}_{IJ} \phi^{ I \prime }  \phi^{ J \prime } +\frac{1}{a^2} f_{X}\mathcal{G}_{IJ} \phi^{I \prime }\partial_{0}\delta \phi^{J} ,
    \\ \nonumber -(\rho+p)v&=&\frac{2}{a^2} f_{R} (\psi^{\prime}+ \mathcal{H} \varphi)+ \frac{1}{a^2} \partial_{0} \delta f_{R} - \frac{\mathcal{H}}{a^2} \delta f_{R} -\frac{\varphi}{a^2} f_{R}^{\prime} + \frac{1}{2a^2} f_{X} \mathcal{G}_{IJ} \partial_{0}\phi^{I} \delta \phi^{J},
    \\ \nonumber  \psi-\varphi &=&\frac{\delta f_{R} }{f_{R}},
     \\ \nonumber \delta p &=& \frac{\mathcal{H}^{\prime} +2\mathcal{H}^2}{a^2} \delta f_{R} + \frac{f_{R}}{a^2} \big[-\psi^{''}+\nabla^{2}\psi - \mathcal{H}(\varphi^{\prime}+5\psi^{\prime})- (2\mathcal{H}^{\prime}+4\mathcal{H}^{2})\varphi \big]  +\frac{2\varphi}{a^2} f_{R}^{''} \\ \nonumber  &&  -\frac{2\varphi}{a^2} \mathcal{H} f_{R}^{\prime}-\frac{4\psi}{a^2} \mathcal{H} f_{R}^{\prime} -\frac{1}{a^2} \partial_{0}^{2} \delta f_{R}+2\frac{\mathcal{H}}{a^2} \partial_{0}\delta f_{R} +\frac{\varphi^{\prime}}{a^2} f_{R}^{\prime} +\frac{\nabla^{2}}{a^2} \delta f_{R},
  \\ \nonumber
  0&=&
-\delta f_{X} \partial_{\eta} \big(a^2 \phi^{L \prime } \big) + f_{X} \partial_{\eta} \big( 2 \varphi a^2 \phi^{L \prime } \big) - f_{X} \partial_{\eta} \big( a^2 (\varphi-3\chi) \phi^{ L \prime } \big) - f_{X} \partial_{\eta} \big( a^2 \partial_{\eta} \delta \phi^{L} \big) \\ \nonumber && +f_{X} \partial_{i} \big(  a^2 \partial_{i} \delta \phi^{L} \big) + a^4 (\varphi -3\psi) \Big[  -\frac{1}{a^2} f_{X} \Gamma^{L}_{KJ} \phi^{K \prime }  \phi^{J \prime } -\frac{1}{a^2} f_{X}^{\prime}\phi^{ L \prime }+\mathcal{G}_{IL} f_{I}   \Big]
 \\ \nonumber && +a^4 \Big[- \frac{1}{a^2} \delta \Gamma^{L}_{KJ} \phi^{ K \prime }  \phi^{J \prime } -  \frac{2}{a^2} f_{X} \Gamma^{L}_{KJ} \partial_{\eta}\delta\phi^{K} \phi^{ J \prime } + \frac{2\varphi}{a^2} f_{X} \Gamma^{L}_{KJ} \phi^{K \prime }  \phi^{J \prime } 
\\ \nonumber &&  - \frac{1}{a^2} \delta f_{X}  \Gamma^{L}_{KJ} \phi^{K \prime }  \phi^{ J \prime } - \frac{1}{a^2} \partial_{\eta} \delta f_{X}  \phi^{L \prime } + \frac{2\varphi}{a^2} f_{X}^{\prime} \phi^{ L \prime } - \frac{1}{a^2} f_{X}^{\prime} \partial_{\eta} \delta\phi^{L} + \mathcal{G}^{IL}_{,K} \delta \phi^{K} f_{I} \\  && + \mathcal{G}^{IL} \delta f_{I} \Big]. \label{PertEq}
\end{eqnarray}
Unlike in general relativity where we have $ \psi=\varphi $, here the non-linearity of $ f(R, \phi^I, X) $ w.r.t the Ricci scalar causes gravitational anisotropy,   $ \psi=\varphi +\frac{\delta f_{R} }{f_{R}} $, which can have interesting implications to be observed. By use of the components (00) and (ij) with $ i \neq j $ from equations \eqref{PertEq}  one can obtain potentials   in Fourier space for small scales
\begin{eqnarray}
\Big(\frac{k}{a} \Big)^2 \psi &\simeq & \frac{1}{2f_{R}} \Big( \Big(\frac{k}{a} \Big)^2 \delta f_{R} - \delta \rho \Big) \nonumber ,
\\ \Big(\frac{k}{a} \Big)^2  \varphi &\simeq & -\frac{1}{2f_{R}} \Big(\Big(\frac{k}{a}\Big)^2 \delta f_{R} + \delta \rho \Big), \label{poi}
\end{eqnarray}
in which we have used $ \delta f \simeq -2 f_R (\frac{k}{a})^2 (2\psi -\varphi) $ and $ \delta \rho $ is the energy density perturbation in comoving gauge. The perturbation in Ricci scalar is
\begin{equation}
\delta R = -2  \Big(\frac{k}{a} \Big)^2 \Big(\varphi +2 \frac{\delta f_{R}}{f_{R}} \Big). \label{deltaR}
\end{equation}
Now let us consider the case $ f_{RX}=0 $. Then equation \eqref{deltaR} results in
\begin{equation}
\delta R= -2\Big(\frac{k}{a}\Big)^2  \frac{\varphi + 2\frac{f_{RI}}{f_{R}} \delta\phi^{I} }{1+4 \big(\frac{k}{a}\big)^2 \frac{f_{RR}}{f_{R}}}. \label{R1}
\end{equation}
The scalar field equations in small scales can be approximated as:
\begin{equation}
-f_{X} \Big(\frac{k}{a}\Big)^2 \delta\phi^{L}+ \mathcal{G}^{IL}_{,K} \delta\phi^{K} + 
\mathcal{G}^{IL} f_{R,L} \delta R\simeq 0. \label{AproSF}
\end{equation}
By substitutin of \eqref{R1} in \eqref{AproSF}, we can  write this equation as $ \alpha^{I}_{\; J} \delta\phi^{J}=\beta^{I} \varphi $,
where:
\begin{equation}
\alpha^{I}_{\; N}:=f_{I} \mathcal{G}^{IL}_{\quad ,N}-f_{X} \Big( \frac{k}{a} \Big)^{2} \delta^{I}_{N} -4\Big( \frac{k}{a} \Big)^{2} \frac{\mathcal{G}^{IL}f_{RI} f_{RN}}{f_{R}+\big( \frac{k}{a} \big)^{2} f_{RR}},
\end{equation}
and
\begin{equation}
\beta^I:=2 \Big( \frac{k}{a} \Big)^2 \frac{ \mathcal{G}^{IL}f_{RL} }{ 1+4 \big( \frac{k}{a} \big)^2 \frac{f_{RR}}{f_{R}} }
\end{equation}
are background quantities.
We assume $ \alpha^{I}_{\; N} $ is an invertible matrix.
Then we can write perturbation in scalar fields as
\begin{equation}
\delta\phi^{J}= (\alpha^{-1})^{I}_{\; J} \beta^{J} \varphi .  \label{phi}
\end{equation}
With substituting equation \eqref{phi} in $ \delta R $, it becomes
\begin{equation}
\delta R=-\frac{2 \big(\frac{k}{a} \big)^{2}}{1+4 \big(\frac{k}{a} \big)^{2} \frac{f_{RR}}{f_{R}}} \Big(1+2\frac{f_{RI}}{f_{R}} (\alpha^{-1})^{I}_{\; J} \beta^{J} \Big) \varphi .
\end{equation}
And finally equation \eqref{poi} leads to the Poisson equation
\begin{equation}
\Big(\frac{k}{a}\Big)^{2} \varphi = \frac{1}{2} G_{eff} \rho \delta ,
\end{equation}
where the effective Newton's constant is
\begin{equation}
G_{eff}=\Big( f_{R} +\frac{1}{2}f_{RI} (\alpha^{-1})^{I}_{\; J} \beta^{J} -f_{RR} \frac{\big(\frac{k}{a}\big)^{2}}{1+4\big(\frac{k}{a}\big)^{2} \frac{f_{RR}}{f_{R}}}\Big(1+2\frac{f_{RI}}{f_{R}} (\alpha^{-1})^{I}_{\; J} \beta^{J} \Big) \Big)^{-1}.
\end{equation}
The energy-momentum conservation in small scales is reduced to
\begin{equation}
\delta^{''}+ \mathcal{H} \delta^{\prime} -  \frac{1}{2} G_{eff} \rho \delta \approx 
0.
\end{equation}
The same calculation can be done in obtaining the equation
\begin{equation}
\Big(\frac{k}{a}\Big)^2 \psi =  - \frac{1}{2} q \delta \rho ,
\end{equation}
where 
\begin{equation}
q=\Big( f_{R} -\frac{1}{2}f_{RI} (\gamma^{-1})^{I}_{\; J} \xi^{J} +f_{RR} \frac{(\frac{k}{a})^{2}}{1+2(\frac{k}{a})^{2} \frac{f_{RR}}{f_{R}}} \Big(1+\frac{f_{RI}}{f_{R}} (\gamma^{-1})^{I}_{\; J} \xi^{J} \Big) \Big)^{-1},
\end{equation}
and
\begin{eqnarray}
\gamma^{\; L}_{K}&:=&\mathcal{G}^{IL}_{\quad ,K} f_{I} -f_{X} \Big( \frac{k}{a} \Big)^2 \delta^{L}_{K} -2  \Big( \frac{k}{a} \Big)^2 \frac{\mathcal{G}^{IL}  f_{RI} f_{RK}}{f_{R}+2 \big( \frac{k}{a} \big)^2 f_{RR}},
\nonumber \\
\xi^{L}&:=& 2  \Big( \frac{k}{a} \Big)^2  \frac{\mathcal{G}^{IL} f_{RI} }{1+2  \big( \frac{k}{a} \big)^2 \frac{f_{RR}}{f_{R}}}.
\end{eqnarray}
One can use, Weak Lensing or Integrated Sachs-Wolf effect with the effective potential, 
$ \Phi_{eff}:=\varphi+\psi $,
 for comparing with observations.
In the case of $ f(R) $ gravity, the effective gravitational constant is reduced to
\begin{equation}
G_{eff}=f_{R}^{-1}\frac{1+4\big( \frac{k}{a} \big)^2 \frac{f_{RR}}{f_{R}}}{1+3\big( \frac{k}{a} \big)^2 \frac{f_{RR}}{f_{R}}}.
\end{equation}
For Brans-Dicke theory with$
f(R,\phi , X)=\phi R+\frac{2 \omega_{BD}}{\phi} \partial_{\mu}\phi \partial^{\mu} \phi
$,
where $ \omega_{BD} $ is Brans-Dicke parameter, we can obtain
\begin{equation}
G_{eff}=\phi^{-1} \frac{4+2\omega_{BD}}{3+2\omega_{BD}}.
\end{equation}
For  non-minimal coupling two field model with the action
\begin{equation}
S=\int d^4 x \sqrt{- g} \Big[ \Big(1+ \frac{1}{2} \zeta_{\phi} \phi^2 +\frac{1}{2} \zeta_{\chi} \chi^2 \Big)R-\frac{1}{2} (\partial \phi)^{2} - \frac{1}{2} e^{2b(\phi)} (\partial \chi)^{2} \Big],
\end{equation}
where $ \zeta_{\phi} $ and $ \zeta_{\chi} $ are dimensionless constant, the metric of field space becomes
\begin{equation}
\mathcal{G}_{ij}=diag(1,e^{2b(\phi)}).
\end{equation}
One can find
\begin{eqnarray}
\alpha^{\phi}_{\; \phi}&=& -\Big( \frac{k}{a} \Big)^2 -4  \Big( \frac{k}{a} \Big)^2 \frac{\zeta_{\phi}^{2} \phi^{2}}{1+\zeta_{\phi} \frac{1}{2} \phi^2 + \zeta_{\chi} \frac{1}{2}\chi^2},
\nonumber \\ \nonumber
 \alpha^{\phi}_{\; \chi}&=&-4  \Big( \frac{k}{a} \Big)^2 \frac{\zeta_{\phi} \zeta_{\chi} \phi \chi}{1+\zeta_{\phi} \frac{1}{2} \phi^2 + \zeta_{\chi} \frac{1}{2}\chi^2},
\\ \nonumber
\alpha^{\chi}_{\; \phi}&=&-4  \Big( \frac{k}{a} \Big)^2 \frac{e^{-2 b(\phi)}\zeta_{\phi} \zeta_{\chi} \phi \chi}{1+\zeta_{\phi} \frac{1}{2} \phi^2 + \zeta_{\chi} \frac{1}{2}\chi^2} -2b_{,\phi} e^{-2 b(\phi)} \zeta_{\phi} \phi R,
\\ \nonumber
 \alpha^{\chi}_{\; \chi}&=&-\Big( \frac{k}{a} \Big)^2 -4  \Big( \frac{k}{a} \Big)^2 e^{-2 b(\phi)} \frac{\zeta_{\chi}^{2} \chi^{2}}{1+\zeta_{\phi} \frac{1}{2} \phi^2 + \zeta_{\chi} \frac{1}{2}\chi^2},
\\ \nonumber
\beta^{\phi}&=&2 \Big( \frac{k}{a} \Big)^2 \zeta_{\phi} \phi,
\\ 
 \beta^{\chi}&=&2 \Big( \frac{k}{a} \Big)^2 e^{-2 b(\phi)}  \zeta_{\chi} \chi,
\end{eqnarray}
and finally, the effective gravitational constant is obtained as
\begin{eqnarray}
G_{eff}^{-1}&=&1+ \frac{1}{2} \zeta_{\phi}  \phi^2 + \frac{1}{2} \zeta_{\chi} \chi^{2} - \Big( \frac{k}{a} \Big)^2  \Big(
 \zeta_{\phi} \phi (\alpha^{-1})^{\; \phi}_{\phi} \zeta_{\phi} \phi
 +\zeta_{\phi} \phi (\alpha^{-1})^{\; \phi}_{\chi} e^{-2 b(\phi)}  \zeta_{\chi} \chi
\nonumber \\
&&+ \zeta_{\chi} \chi (\alpha^{-1})^{\; \chi}_{\phi} \zeta_{\phi} \phi
 + \zeta_{\chi} \chi (\alpha^{-1})^{\; \chi}_{\chi} e^{-2 b(\phi)}  \zeta_{\chi} \chi
\Big).
\end{eqnarray}
\section{Conformal transformation}
\label{conformal transformation}
The equivalence between $ f(R) $  and scalar-tensor theories is well known (see reference \cite{15}). 
In last two  sections we studied the general form of the action with multiple fields, in this section we focus on  conformal transformation between Jordan and Einstein frames that are special cases of the action \eqref{action} 
and finally obtain their field equations. 
The D-dimensional action in Jordan frame is given by
\begin{equation}
S_{J}=\int d^{D}x \sqrt{-g} [\frac{1}{2}f(\phi^{i},R)-\frac{1}{2}\mathcal{G}_{ij} \partial_
{\mu} \phi^{i} \partial^{\mu} \phi^{j} -V(\phi^{i}) ], \label{PriAction}
\end{equation} 
in which we have used $ i,j,k,...=1,2,3,...,N $ to label scalar fields in this frame, a general  potential is represented by
 $ V(\phi^{i}) $,   $ f(\phi^{i},R) $ is a function of scalar fields and Ricci scalar, and contains their coupling.
  In reference \cite{16}  the case of $ f(\phi^{i} , R) = \Big( M_{pl}^2 + \sum_{i}  \xi_{i} ( \phi^{i} )^{2} \Big) R $ for two field in  four-dimensional have been studied.
The generalized Klein-Gordon equation becomes
\begin{equation}
\square \phi^{i}+\Gamma^{i}_{kj} \nabla_{\mu}\phi^{k} \nabla^{\mu}\phi^{j}+ \mathcal{G}^{il} (\frac{f}{2}-V)_{,l}=0. \label{CE}
\end{equation} 
Here we define the Lagrangian of scalar fields as follow
\begin{equation}
\mathcal{L}_{\phi} :=-\frac{1}{2}\mathcal{G}_{ij} \partial_
{\mu} \phi^{i} \partial^{\mu} \phi^{j} -V(\phi^{i}),
\end{equation}
with this definition and by use of equation \eqref{CE}, scalar field energy-momentum tensor,
\begin{equation}
T^{(\phi)}_{\mu \nu}:=-\frac{2}{\sqrt{-g}} \frac{\delta (\sqrt{-g} \mathcal{L}_{\phi})}{\delta(g^{\mu \nu})},
\end{equation} 
satisfy the relation
\begin{equation}
\nabla^{\mu} T^{(\phi)}_{\mu \nu}=-\frac{1}{2} \frac{\partial f(\phi^{i},R)}{\partial \phi^{j}} \nabla_{\nu} \phi^{j}.
\end{equation}
After some simple calculation conservation of matter energy-momentum tensor becomes
\begin{equation}
\nabla^{\mu} T^{(M)}_{\mu \nu}=0.
\end{equation}
The action \eqref{PriAction} by use of auxiliary scalar field $ \chi $ can be written as
\begin{equation}
S_{J}=\int d^{D}x \sqrt{-g} [\frac{1}{2}f_{,\chi}(\phi^{i},\chi)R-\frac{1}{2}\mathcal{G}_{ij} 
\partial_{\mu} \phi^{i} \partial^{\mu} \phi^{j} -U(\phi^{i},\chi) ], \label{AxAction}
\end{equation}
where $2 U(\phi^{i},\chi):=2 V(\phi^{i})-f(\phi^{i},\chi)+f_{,\chi}(\phi^{i},\chi)
\chi$ is a new effective potential. The variation of the action \eqref{AxAction} w.r.t auxiliary field yields
\begin{equation}
f_{,\chi \chi}(\phi^{i},\chi) (R-\chi)=0.
\end{equation}
If we consider the case of $ f_{,\chi \chi}\neq 0 $ and $ R-\chi=0 $,  the action  \eqref{AxAction} leads to the action \eqref{PriAction} in Jordan frame. The condition $ f_{,\chi \chi}\neq 0 $ implies that new scalar degree of freedom comes from non-linearity nature of  $ f(\phi^{i},R) $ respect to the Ricci scalar.
This action looks like $ N+1 $ scalars with non-minimal coupling. By use of a convenient conformal transformation, the non-minimal coupling can be removed. 
The metrics in two frames are related by 
\begin{equation}
\hat{g}_{\mu \nu}= \Omega^{2}(x) g_{\mu \nu}, \qquad \sqrt{- \hat{g}}=\Omega^{D}(x) \sqrt{-g}, \label{metric}
\end{equation}
where we have used $ \Omega^{2} > 0 $ to preserve casual structure.
With this transformation the Ricci scalar becomes 
\begin{equation}
\hat{R}=\Omega^{2} \Big[  R-\frac{2(D-1)}{\Omega} \square\Omega -(D-1)(D-4) \Omega^{-2}  \nabla^{\mu}\Omega \nabla_{\mu} \Omega \Big],
\end{equation}
where $ \square\Omega=\frac{1}{\sqrt{-g}} \partial_{\mu} (\sqrt{-g}\partial^{\mu} \Omega) $. The first term in \eqref{AxAction} becomes
\begin{eqnarray}
\int d^D x \frac{\sqrt{-g}}{2} f_{,\chi}(\phi^i ,\chi)R &=& \int d^D x \frac{\sqrt{-\hat{g}}}{2} \Omega^{-D} f_{,\chi}(\phi^{i},\chi) \Big(  \Omega^2 \hat{R}+2(D-1)\Omega^{-1} \square\Omega  \nonumber \\ 
&&+(D-1)(D-4) \Omega^{-2}  \nabla^{\mu}\Omega \nabla_{\mu} \Omega  \Big).
\end{eqnarray}
By the choice of
\begin{equation}
\Omega^{D-2}(x)=\frac{1}{M_{(D)}^{D-2}} f_{,\chi}(\phi^{i},\chi),
\end{equation} 
where $ M_{(D)}^{D-2} $ is reduced Planck mass in D-dimension  (in four-dimension $ M_{(4)}=2.43 \times 10^{18} GeV $), we can obtain the action in new frame
\begin{equation}
S_{E}=\int d^{D}x \sqrt{-\hat{g}} \Big[\frac{M_{(D)}^{D-2}}{2} \hat{R}-\frac{1}{2} 
\hat{\mathcal{G}}_{IJ} \hat{g}^{\mu \nu}\partial_{\mu} \phi^{I} \partial_{\nu} 
\phi^{J} -\hat{U}(\phi^{I}) \Big],
\end{equation} 
where
\begin{equation}
\hat{U}(\phi^{I}):=U(\phi^i , \chi) \Bigg( \frac{M_{(D)}^{D-2}}{f_{,\chi}} \Bigg)^{\frac{D}
{D-2}}
\end{equation}
is the effective potential in Einstein frame,
$ \{\phi^I\}:=\{\phi^i, \chi \} $ and $ I,J,K,...=1,2,3,...(N+1) $,
and the metric in new frame is modified to
\begin{eqnarray}
\hat{\mathcal{G}}_{ij}&=&\frac{M_{(D)}^{D-2}}{f_{,\chi}} \mathcal{G}_{ij}+ \Big( \frac{D-1}{D-2}\Big)  M_{(D)}^{D-2} \frac{f_{,\chi i} f_{,\chi j}}{f_{,\chi}^{2}},
\nonumber \\ \nonumber
\hat{\mathcal{G}}_{\chi i}&=&\hat{\mathcal{G}}_{i \chi}= \Big( \frac{D-1}{D-2}\Big) M_{(D)}^
{D-2} \frac{f_{,\chi i} f_{,\chi \chi}}{f_{,\chi}^{2}},
\\ 
\hat{\mathcal{G}}_{\chi \chi}&=&\Big( \frac{D-1}{D-2}\Big)  M_{(D)}^{D-2} \Big(\frac{f_{,\chi 
\chi}} {f_{,\chi }}\Big)^{2}.
\end{eqnarray}
The field equation in new frame contains $ N+1 $ second order differential equations but in old frame  the higher order derivatives is appeared which can be solved more difficultly. Some simple cases are;
\begin{itemize}
\item 
In the case of minimal coupling between $ f(R) $ and scalar fields, $ f_{\chi i}=0 $ leads to  simple field space metric,
\begin{eqnarray}
\hat{\mathcal{G}}_{ij} &=&\frac{M_{(D)}^{D-2}}{f_{,\chi}} \mathcal{G}_{ij},
\nonumber \\ \nonumber
\hat{\mathcal{G}}_{\chi i}&=&\hat{\mathcal{G}}_{i \chi}=0,
\\ 
\hat{\mathcal{G}}_{\chi \chi}&=&\Big( \frac{D-1}{D-2}\Big)  M_{(D)}^{D-2} \Big(\frac{f_{,\chi \chi}} {f_{,\chi }} \Big)^{2}.
\end{eqnarray}
In this case, we can define new field $ \psi:= M_{(D)}^{\frac{D-2}{2}} \sqrt{\frac{D-1}{D-2}} \ln f_{\chi}$, thence kinetic terms become
\begin{equation}
-\frac{1}{2}  \hat{\nabla}_{\mu}\psi \hat{\nabla}^{\mu}\psi -\frac{1}{2} \mathcal{G}_{ij} e^{2b(\psi)} \hat{\nabla}_{\mu}\phi^{i} \hat{\nabla}^{\mu}\phi^{j},
\end{equation}
where  we have defined
\begin{equation}
b(\psi):=\frac{1}{2} \ln \big( M_{(D)}^{D-2} \big) - \sqrt{\frac{D-2}{D-1}} \frac{1}{2  M_{(D)}^{\frac{D-2}{D}} } \psi .
\end{equation} 
This case for $ f(R) $ with single field model have been studied in \cite{17},
 in Jordan frame becomes a two-fields model with non-canonical kinetic term. 
\item
In the case of $ f(R, \phi^{i})=h(\phi^{i}) R+P(\phi^{i}, X) $, where $ h(\phi^{i}) $ is a function of scalar fields  and  $ P(\phi^{i}, X) $ is a function of both scalar fields and kinetic terms, we can see $ \hat{\mathcal{G}}_{i\chi} = 0 = \hat{\mathcal{G}}_{\chi \chi} $ and  $ \hat{\mathcal{G}}_{ij} = \frac{M_{(D)}^{D-2}}{h^2} \Big(  \mathcal{G}_{ij} h+\frac{D-1}{D-2} h_i h_j \Big)  $ where in four dimensional spacetime is reduced to
\begin{equation}
\hat{\mathcal{G}}_{ij} = \frac{M_{pl}^2}{h^2} \bigg(  \mathcal{G}_{ij} h + \frac{3}{2} h_i h_j \bigg).
 \end{equation}
This result is the same as that of reference \cite{12}.

\end{itemize}
In conformal transformation, we relabel the metric. At the classical level, frames are mathematically equivalent and all observations should be same \cite{18}.
Till now, we have considered the gravitation part only. In presence of matter action, the conformal transformation leads to coupling of the scalar fields to matter part, which introduce new properties. For example leads to the $ x^{\mu} $ dependence of particle masses. We can write the matter action in Jordan frame as 
\begin{equation}
S_{J}^{(M)}=\int d^D x \mathcal{L}_{M}(g_{\mu \nu},\Psi_{M})=\int d^D x \mathcal{L}_{M} \Big( \frac{\hat{g}_{\mu \nu}}{\Omega^2},\Psi_{M} \Big).
\end{equation}
With the definition of energy-momentum tensor we can obtain
\begin{equation}
\hat{T}^{(M)}_{\mu \nu}:=\frac{-2}{\sqrt{-\hat{g}}} \frac{\delta  \mathcal{L}}{\delta \hat{g}^{\mu \nu}}=T^{(M)}_{\mu \nu} \Bigg( \frac{M_{(D)}^{D-2}}{f_{,\chi}} \Bigg). \label{MatterTensor}
\end{equation}
Also derivation of matter Lagrangian w.r.t scalar fields yields 
\begin{equation}
\frac{\partial \mathcal{L}_{M}}{\partial \phi^{I}}= \kappa_{(D)} \sqrt{-\hat{g}} Q_{I} \hat{T}^{(M)},
\end{equation}
where $  \kappa_{(D)}:=(M_{(D)}^{D-2})^{-1} $, $ \hat{T}^{(M)} $ is the trace of matter energy-momentum tensor,
$ Q_{I}:=- M_{(D)}^{D-2} \frac{ \Xi_{,I}}{2 \Xi} $ and $ \Xi:=\Omega^2 $.
The Klein-Gordon equation in Einstein frame without any matter fields can be written as (here we do not use hat notation)
\begin{equation}
\square \phi^{I}+\Gamma^{I}_{KJ} g^{\alpha \beta} \partial_{\alpha}\phi^{J} \partial_{\beta}\phi^{K}-\mathcal{G}^{IM} U_{M}=0. \label{CJ}
\end{equation}
In presence of matter action, this equation becomes
\begin{equation}
 \square \phi^{I}+\Gamma^{I}_{KJ} g^{\alpha \beta} \partial_{\alpha}\phi^{J} \partial_{\beta}\phi^{K}-\mathcal{G}^{IM} U_{M}+\kappa_{(D)} \mathcal{G}^{IJ} Q_{I} T^{(M)}=0. \label{CJM}
\end{equation}
The field equations \eqref{CE} and \eqref{CJ} in four-dimensional FLRW spacetime become, respectively
\begin{eqnarray}
 \phi^{i ''}+2\mathcal{H} \phi^{i \prime} + \Gamma^{i}_{kj} \phi^{j \prime} \phi^{k \prime} -a^2  \mathcal{G}^{il} (\frac{f}{2}-V)_{,l} &=&0,
 \nonumber \\
  \phi^{I ''}+2\mathcal{H} \phi^{I \prime} +\Gamma^{I}_{KJ}  \phi^{J \prime} \phi^{K \prime}+a^2 \mathcal{G}^{IL} U_{L}&=&0,
\end{eqnarray}
or in compact form
\begin{eqnarray}
\mathcal{D}_{\eta}(a^2 \phi^{\prime i}) &=&  a^4 \mathcal{G}^{il} (\frac{f}{2}-V)_{,l},
\nonumber \\
\mathcal{D}_{\eta}(a^2 \phi^{I \prime}) &=& a^4 \mathcal{G}^{IL} U_{L}.
\end{eqnarray}
Also equation \eqref{CJM} is reduced to
\begin{equation}
\phi^{I ''}+\Gamma^{I}_{KJ}  \phi^{J \prime} \phi^{K \prime}+2\mathcal{H} \phi^{I \prime} +a^2 \mathcal{G}^{IL} U_{L}-a^2 \kappa_{(4)} \mathcal{G}^{IJ} Q_{I} T^{(M)}=0.
\end{equation}
In Einstein frame one can separate scalar field part of the action,
\begin{equation}
S_{\phi}:=\int d^4 x \sqrt{-g} \Big[-\frac{1}{2} \mathcal{G}_{IJ} \partial_{\mu} \phi^{I} \partial^{\mu} \phi^{J}-U(\phi^{I}) \Big] \label{ScalarAction}.
\end{equation}
The energy-momentum, energy density and pressure for action \eqref{ScalarAction} become
\begin{eqnarray}
T^{(\phi) \mu}_{\nu} &=& \mathcal{G}_{IJ} \partial^{\mu} \phi^{I}\partial_{\nu} \phi^{J} - \delta_{\nu}^{\mu} \Big(\frac{1}{2}\mathcal{G}_{IJ} \partial_{\alpha} \phi^{I} \partial^{\alpha} \phi^{J} +U(\phi^{I})\Big),
\nonumber \\ \nonumber
-T^{(\phi) 0}_{0} &=&\rho_{(\phi)} = \frac{1}{2a^2} \mathcal{G}_{IJ} \phi^{I \prime} \phi^{J \prime} +U(\phi^{I}),
\\ 
\frac{1}{3} T^{(\phi) k}_{k} &=& p_{(\phi)} = \frac{1}{2a^2} \mathcal{G}_{IJ} \phi^{I \prime} \phi^{J \prime} - U(\phi^{I}).
\end{eqnarray}
It can be seen with simple manipulation that the conservation  equations of energy for scalar fields and matter are
\begin{eqnarray}
\rho^{\prime}_{(\phi)}+3 \mathcal{H} (\rho_{(\phi)}+p_{(\phi)}) &=& -\kappa_{(4)} Q_{I} \phi^{ I \prime} T^{(M)} ,
\nonumber \\ 
\rho^{\prime}_{M}+3 \mathcal{H} (\rho_M+p_M) &=& +\kappa_{(4)} Q_{I}  \phi^{I \prime} T^{(M)},
\end{eqnarray}
and in covariant form
\begin{eqnarray}
\nabla_{\mu} T^{(\phi) \mu}_{\nu}&=&-\kappa_{(4)} Q_{I} \nabla_{\nu} \phi^{I} T^{(M)} ,
\nonumber \\ 
\nabla_{\mu} T^{(M) \mu}_{\nu}&=&+\kappa_{(4)} Q_{I} \nabla_{\nu} \phi^{I} T^{(M)} .
\end{eqnarray}
Thus the energy-momentum conservation in Einstein frame can be written as $ \nabla_{\mu} \Big(T^{(\phi) \mu}_{\nu} +T^{(M) \mu}_{\nu} \Big)=0 $.

In  general form, metric perturbations can be written as $ g_{\mu \nu}=\bar{g}_{\mu \nu}+\delta g_{\mu \nu} $.
We can write the conformal factor as
\begin{equation}
\Xi(\eta ,x^{i})=\bar{\Xi}(\eta) \Big(  1+\frac{\delta \Xi (\eta ,x^{i})}{\bar{\Xi}(\eta)}  \Big). 
\end{equation}
Therefore, perturbation $ \delta g_{\mu \nu} $ transform as
\begin{equation}
\delta \hat{ g}_{\mu \nu}= \bar{\Xi} \delta g_{\mu \nu} + \bar{g}_{\mu \nu} \delta \Xi .
\end{equation}
Thus we can see for the components of metric, that are vanished on background level, the corresponding perturbations are independent of 
conformal factor perturbation, and specially   in the case of flat FLRW these are conformal invariant.
The conformal transformation \eqref{metric} for FLRW metric in background level leads to (we use hat notation for the Einstein frame)
\begin{equation}
d\hat{t}=\sqrt{\Xi} dt,
\quad \hat{a}(\hat{t})=\sqrt{\Xi} a(t),
\quad d\hat{x}^{i}=dx^{i},
\quad d\hat{\eta}=d\eta .
\end{equation}
One can obtain the comoving Hubble parameter,
\begin{equation}
\hat{\mathcal{H}}=\frac{\Xi^{\prime}}{2 \Xi}+\mathcal{H}.
\end{equation}
We can see some properties, for example the Hubble crossing,
are not equivalent in two frames. 
The general form of perturbations can be  classified according to their helicity  \cite{19} (scalar, vector and tensor modes) by
\begin{equation}
ds^2=a^2(\eta) \big[  -(1+2\varphi)d\eta^{2}-2(B_i +\beta_{,i}) d\eta dx^{i}+[(1-2\psi) \delta_{ij}+2\partial_{i} \partial_{j} E + 2\partial_{(i}C_{j)} + E_{ij}] dx^{i} dx^{j} \big],
\end{equation} 
where $ \varphi $, $ \beta $, $ \psi $ and $ E $ are scalar perturbations, $ B_i $ and $ C_{j} $ are transverse vector perturbations and $ E_{ij} $ is  traceless and transverse tensor perturbation.
The metric perturbation quantities obtain as
\begin{equation}
\hat{\varphi}=\varphi + \frac{\delta \Xi}{2 \bar{\Xi}}, 
\quad \hat{\beta}=\beta ,
\quad \hat{\psi}=\psi -  \frac{\delta \Xi}{2 \bar{\Xi}}, 
\quad \hat{E}=E, \label{ConformalMetPer}
\end{equation} 
the vector and tensor perturbations are conformal invariant.
The relations \eqref{ConformalMetPer} show  the Newtonian gauge in one frame is transformed into the same gauge in the new frame, but other gauges like synchronous gauge, is not equivalent in two frames and lead to the additional redundant degree of freedom in new frame \cite{20}.
If we take  a gauge with the condition $ \delta\Xi=0 $, the metric perturbations are conformal invariant. From the equation \eqref{MatterTensor}, the energy-momentum tensor in background and first order of perturbations in D-dimensional  become, respectively
\begin{eqnarray}
\bar{\Xi}^{\frac{D}{2}} \hat{T}^{(M) \mu}_{\nu} &=& T^{(M)\mu}_{\nu} ,
\nonumber \\
\bar{\Xi}^{\frac{D}{2}} \delta \hat{T}^{(M) \mu}_{\nu} &=& \delta T^{(M)\mu}_{\nu} -\frac{D}{2} T^{(M) \mu}_{\nu} \frac{\delta \Xi}{\bar{\Xi}}, \label{FF}
\end{eqnarray} 
where 
\begin{equation}
\frac{\delta \Xi}{\bar{\Xi}} =\frac{2}{D-2} \frac{\delta f_{,\chi}}{f_{,\chi}}.
\end{equation}
Thus with the condition $\delta \Xi=0 $, the quantity $ \frac{\delta T^{(M) \mu}_{\nu}}{T^{(M) \alpha}_{\beta}} $  is conformal invariant.
In single field case the uniform conformal transformation slicing gauge, $\delta \Xi=0 $, coincide perfectly with the comoving and constant field gauges, $ \delta \Xi \propto \delta \phi$.\footnote{We have used the comoving curvature perturbation as the curvature perturbation on comoving slice of effective fluid by $ \delta T^{0}_{i}=0 $.} But in general case, the uniform conformal transformation slicing gauge is not equivalent to the comoving or constant field gauges.
For constant $ Q_I $  one can integrate from $ Q_I $ and obtain
\begin{equation}
\Xi=e^{-2  \kappa_{(D)}  Q_{I} \phi^{I}},
\end{equation}
and the equation \eqref{FF} and condition $ \delta\Xi=0 $  are reduced to
\begin{eqnarray} 
e^{-  \kappa_{(D)}  Q_{I} \phi^{I}} \delta \hat{T}^{(M) \mu}_{\nu} &=& \delta T^{(M)\mu}_{\nu} - \kappa_{(D)} D T^{(M) \mu}_{\nu}  Q_{I} \delta\phi^{I},
\nonumber \\
Q_{I} \delta\phi^{I} &=& 0.
\end{eqnarray} 
\section{Conclusion}

In this work we have studied the scalar perturbations in Newtonian gauge for the general form of  Multiple non-canonical scalar fields with non-minimal coupling. In small scales, Newton's constant of gravitation is modified for such multiple degree of freedom models  and explicitly in some models  is obtained. Finally equations of motion in Jordan and Einstein frames in presence of matter  have been obtained and conservation of energy-momentum tensor is studied. The conformal transformation leads to coupling   scalar fields to matter part. With assuming the conservation of energy-momentum tensor in Jordan frame, we can see this conservation in Einstein frame is not valid only for matter action but it needs adding an  energy-momentum tensor of   scalar fields. 
\acknowledgments
H. A. would like to thank David I. Kaiser  for  helpful discussions.


\begin{thebibliography}{99}
\bibitem{1} S. Perlmutter, G. Aldering, G. Goldhaber, R. A. Knop, et
al., \emph{Measurements of Omega and Lambda from 42 high redshift supernovae}, \emph{Astrophys. J.}  {\bf 517} (1999) 565-586 [astro-ph/9812133].
\bibitem{2} B. Ratra, P. J. E. Peebles, \emph{Cosmological Consequences of a Rolling Homogeneous Scalar Field}, \emph{Phys. Rev. D} {\bf 37} (1988) 3406.
\bibitem{3}
I. Zlatev, L. M. Wang, P. J. Steinhardt, \emph{Quintessence, cosmic coincidence, and the cosmological constant}, \emph{Phys. Rev. Lett.} {\bf 82} (1999) 896-899.
\bibitem{4b} L. Amendola, \emph{Scaling solutions in general nonminimal coupling theories}, \emph{Phys. Rev. D} {\bf 60} (1999)043501 [astro-ph/9904120].
\bibitem{5} T. P. Sotiriou, V. Faraoni, \emph{f(R) Theories Of Gravity}, \emph{Rev. Mod. Phys.}  {\bf 82} (2010) 451 [gr-qc/0805.1726].
\bibitem{6} C. Brans, R. H. Dicke, \emph{Mach's Principle and a Relativistic Theory of Gravitation}, \emph{Phys. Rev.} {\bf 124 } (1961) 925-935.
\bibitem{7} J. D. Bekenstein, \emph{Relativistic gravitation theory for the MOND paradigm}, \emph{Phys. Rev. D} {\bf 70} (2004) 083509 [astro-ph/040369].
\bibitem{8} P. D. Mannheim, \emph{Alternatives to dark matter and dark energy}, \emph{Prog. Part. Nucl. Phys.} {\bf 56} (2006) 340-445 [astro-ph/0505266].
\bibitem{9} D. Lovelock, \emph{The Einstein tensor and its generalizations},  \emph{J.Math.Phys.} {\bf 12} (1971) 498-501.
\bibitem{10} P. Horava, \emph{Quantum Gravity at a Lifshitz Point}, \emph{Phys.Rev. D} {\bf 79} (2009) 084008  [hep-th/0901.3775].
\bibitem{11} E.V. Linder, \emph{Einstein's Other Gravity and the Acceleration of the Universe}, \emph{Phys. Rev. D} {\bf  81} (2010) 127301, \emph{Phys. Rev.  D} {\bf 82}  (2010) 109902 [astro-ph/1005.3039].
\bibitem{4} N. D. Birrell and P. C. W. Davies, \emph{Quantum Fields in Curved Space}, Cambridge University Press, New York, (1982).
\bibitem{12} D. I. Kaiser, \emph{Conformal Transformations with Multiple Scalar Fields}, \emph{ Phys.Rev. D}  {\bf 81}  (2010) 084044   [gr-qc/1003.1159v2].

\bibitem{Ellis} G. F. R. Ellis, R. Maartens, M. A. H. MacCallum, \emph{Relativistic Cosmology}, Cambridge University Press, (2012).
\bibitem{14} S. Tsujikawa, \emph{Matter density perturbations and effective gravitational constant in modified gravity models of dark energy}, \emph{Phys. Rev. D} {\bf 76} (2007) 023514 [astro-ph/0705.1032 ].
.
\bibitem{15} T. Futamase, K. I. Maeda, \emph{Chaotic Inflationary Scenario in Models Having Nonminimal Coupling With Curvature}, \emph{Phys. Rev. D} {\bf 39} (1989) 399.

\bibitem{16} D. I. Kaiser, E. I. Sfakianakis, \emph{Multifield Inflation after Planck: The Case for Nonminimal Couplings }, \emph{Phys. Rev. Lett.} {\bf 112} (2014) 011302 [astro-ph/1304.0363v3].
\bibitem{17} T. Qiu, J. Q. Xia, \emph{Perturbations of Single-field Inflation in Modified Gravity Theory}, [astro-ph/1406.5902v1]. 
\bibitem{18} N. Deruelle, M. Sasaki, \emph{Conformal equivalence in classical gravity: the example of veiled General Relativity}, \emph{Springer Proc. Phys.} {\bf 137} (2011) 247-260
[gr-qc/1007.3563].
\bibitem{19} D. H. Lyth, A. R. Liddle, \emph{The primordial density perturbation}, Cambridge University Press, New York, (2009).
\bibitem{20} I. A. Brown, A. Hammami, \emph{Gauge Issue in Extended Gravity and $ f(R) $ Cosmology}, \emph{JCAP} {\bf 04} (2012) 002 [gr-qc/1112.0575].
 \end{thebibliography}
\end{document}